\documentclass[aps,pra,twocolumn,superscriptaddress,floatfix,
nofootinbib,showpacs,longbibliography]{revtex4-1}

\usepackage[utf8]{inputenc}  
\usepackage[T1]{fontenc}     
\usepackage[british]{babel}  
\usepackage[sc,osf]{mathpazo}
\usepackage[scaled=0.86]{berasans}  
\usepackage[colorlinks=true, citecolor=blue, urlcolor=blue]{hyperref}  
\usepackage{graphicx} 
\usepackage[babel]{microtype}  
\usepackage{amsmath,amssymb,amsthm,bm,amsfonts,mathrsfs,bbm} 

\usepackage{xspace}  
\usepackage{pgf,tikz}
\usepackage{xcolor}
\usepackage{multirow}
\usepackage{array}
\usepackage{bigstrut}
\usepackage{braket}
\usepackage{color}
\usepackage{natbib}
\usepackage{multirow}
\usepackage{mathtools}
\usepackage{float}
\usepackage{xcolor,colortbl}
\usepackage{color}
\usepackage[justification=justified, format=plain]{subcaption}
\usepackage[justification=raggedright]{caption}
\newcommand{\Tr}{\operatorname{Tr}}

\newcommand{\be}{\begin{equation}}
\newcommand{\ee}{\end{equation}}
\newcommand{\ba}{\begin{eqnarray}}
\newcommand{\ea}{\end{eqnarray}}

\newcommand{\C}{\mathbb{C}}

\newcommand{\spc}[1]{\mathcal{#1}}

\def\>{\rangle}
\def\<{\langle}

\newcommand{\st}[1]{\mathbf{#1}}

\newcommand{\map}[1]{\mathcal{#1}}

\newcommand{\St}{{\mathsf{St}}}

\newtheorem{theo}{Theorem}

\newtheorem{prop}{Proposition}

\begin{document}

\title{Random-Receiver Quantum Communication}	
	
\author{Some Sankar Bhattacharya}
\affiliation{Department of Computer Science, The University of Hong Kong, Pokfulam Road, Hong Kong.}

\author{Ananda G. Maity}
\affiliation{S.N. Bose National Center for Basic Sciences, Block JD, Sector III, Salt Lake, Kolkata 700098, India.}

\author{Tamal Guha}
\affiliation{Physics and Applied Mathematics Unit, Indian Statistical Institute, 203 B.T. Road, Kolkata-700108, India.}

\author{Giulio Chiribella}
\affiliation{Department of Computer Science, University of Oxford, Wolfson Building, Parks Road, United Kingdom.}
\affiliation{Department of Computer Science, The University of Hong Kong, Pokfulam Road, Hong Kong.}

\author{Manik Banik}
\affiliation{School of Physics, IISER Thiruvanathapuram, Vithura, Kerala 695551, India.}

\begin{abstract}
We introduce the task of random-receiver   quantum communication, in which a sender  transmits a quantum message to a receiver chosen from a list of $n$ spatially separated parties. The choice of  receiver is unknown to the sender, but  is  known by the $n$ parties, who coordinate their actions by exchanging classical messages.   In normal conditions,  random-receiver quantum communication requires a noiseless quantum communication channel from the sender to each of the $n$ receivers.  In contrast, we show that random-receiver quantum communication can take place through entanglement-breaking channels if the order of  such channels is controlled by a quantum bit that is accessible through quantum measurements.  Notably, this phenomenon cannot be mimicked by allowing free quantum communication between the sender and any subset of $k<n$ parties.  
\end{abstract}



\maketitle
{\it Introduction:--}  The transmission of quantum messages from a sender to a receiver is the cornerstone of quantum communication. When the identity of the receiver is known,  this task can be achieved with  a  reliable quantum communication channel between the sender and the receiver.   But what if the identity of the receiver is unknown?    

\begin{figure}
	\includegraphics[scale=0.31]{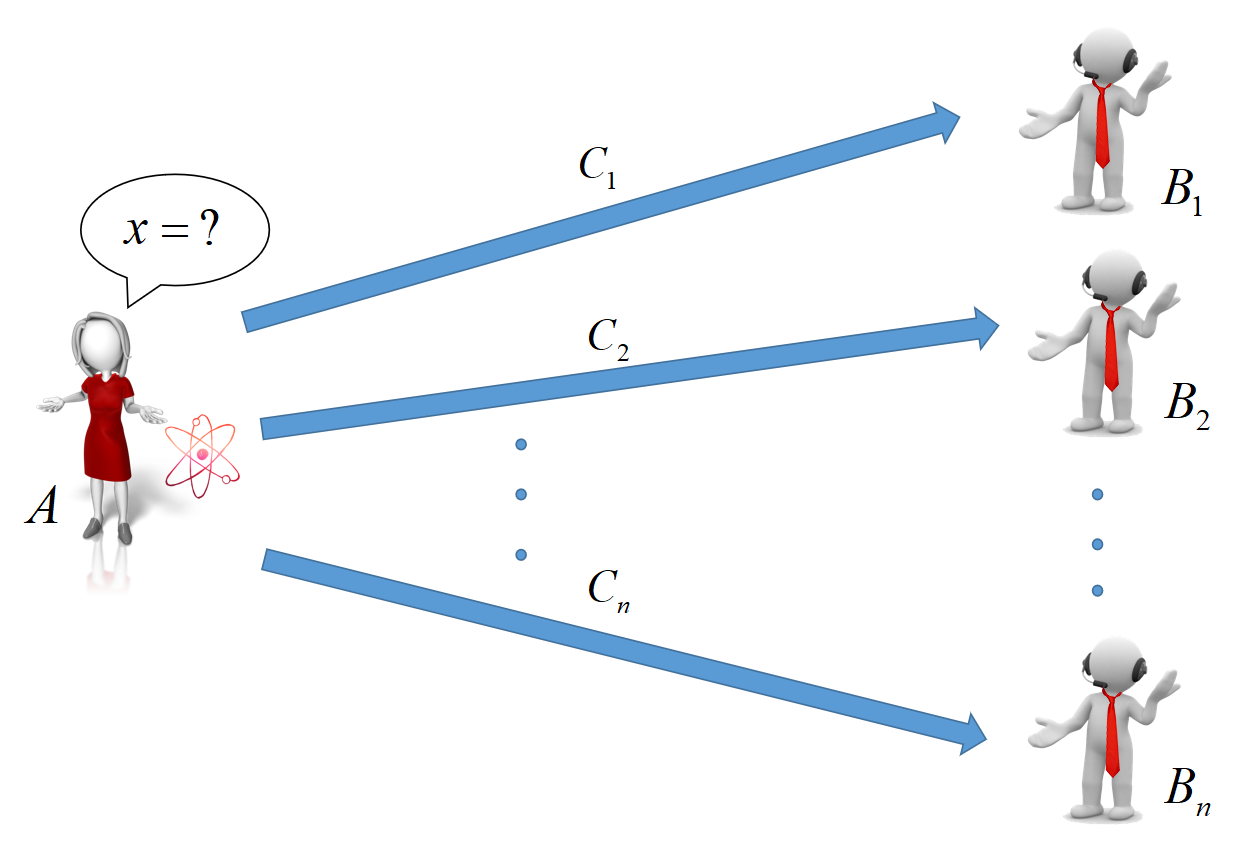}
	\caption{Random-receiver quantum communication task. The sender A wants to transmit a quantum message to one of the n spatially separated receivers $\{B_i\}_{i=1}^n$. The identity of the targeted receiver $x$ is unknown to the sender but known to the receivers. Receivers are allowed to collaborate through classical messages, but quantum communication among them is forbidden.}
	\label{fig:basic}
\end{figure}

Here we introduce a  scenario, which we call {\em random-receiver quantum communication}:  a sender $A$ is connected to $n$ spatially separated parties $(B_i)_{i=1}^n$ through $n$ communication channels $(\map C_i)_{i=1}^n$, as in Figure \ref{fig:basic}. The sender wants to transmit a quantum message to the  $x$-th party, for some $x\in  \{1,\dots, n\}$.  However, the identity of such party ({\em i.e.} the value of $x$)  is unknown to the sender.   This scenario could arise, for example, in  delegated quantum computation, where a client sends inputs to a server, asking it to perform a desired quantum computation on them.  Here the receiver $B_x$ could be one of $n$ servers, and the sender may not know in advance which  server is available to perform the desired computation. In this situation, the sender has to delocalize the message, and send it to all the $n$ servers,  in such a way that the available one can retrieve the message and operate on it. In the following, we will assume that the $n$ parties know the value of $x$ (for example, because they have communicated classically among each other) and cooperate in order to let the message reach party $B_x$.  To coordinate their actions, the parties are allowed to exchange classical messages. We say that a communication protocol is successful if it works for all values of $x \in  \{1,\dots,  n\}$. 

Random-receiver quantum communication is  related to the task of quantum summoning \cite{Kent,Kent12}, where a quantum message has to be revealed at a given set of spacetime points. The crucial difference, however, is that  summoning includes limits on the exchange of signals among the $n$ parties induced by the structure of the underlying spacetime.  In random-receiver quantum communication, classical communication among the $n$ parties is permitted, while quantum communication is forbidden.

 To introduce the task of random-receiver quantum communication, we first consider the simple scenario where the quantum message is a generic state of a quantum bit (qubit), and all the channels from the sender to the receivers are noiseless.  To transmit the quantum state  $|\psi\>  =  \alpha\,  |0\>  +  \beta \,  |1\>$, the sender can encode it in the generalized Greenberger-Horne-Zeilinger (GHZ) state $|\psi_n\>   :  = \alpha\,  |0\>^{\otimes n}  +  \beta  \,  |1\>^{\otimes n}$ and send such state to the $n$ receivers.   To let party $B_x$ retrieve  the message, each of the other $(n-1)$ parties performs a measurement on the Fourier basis $\{  |+\>,  |-\>\}$, $|\pm\>  :  =  (|0\>  \pm |1\>)/\sqrt 2$, collapsing the state of party $B_x$ to $|\psi_s\>   :=   \alpha |0\>  +  (-1)^{s} \beta \,  |1\>$, where $s:= \sum_{y\not =  x}  o_y$ is the sum of the measurement outcomes,   $o_y$ being  the measurement outcome obtained by the $y$-th party.   
Finally, the  $n-1$ parties communicate their outcomes to $B_x$, who  performs the  correction operation $Z^{s}$,  with    $Z  :=  |0\>\<0|  -  |1\>\<1|$.   It is easy to see that party $B_x$ eventually receives the quantum state $|\psi\>$ without any error. All together, this protocol requires $1$ qubit of quantum communication from the sender to each receiver.

Now, suppose that the quantum channels are noisy. For  protocols involving a single round of classical communication to the chosen receiver, we show that perfect random-receiver quantum communication is possible only if each the channels $(\map C_i)_{i=1}^n$ can transfer at least one qubit without errors. This result implies that the simple noiseless protocol presented above is optimal in terms of quantum communication.  Moreover, the result shows that random-receiver quantum communication cannot take place if the channels $(\map C_i)_{i=1}^n$ are {\em entanglement-breaking}.   As it turns out, this impossibility of random-receiver quantum communication with entanglement-breaking channels  holds not only for one-way protocols, but also for protocols involving arbitrarily many rounds of  local operations and classical communication (LOCC)  (see Appendix.\ref{A2}). 

\begin{figure}
	\begin{subfigure}{\linewidth}
		\includegraphics[scale=0.4]{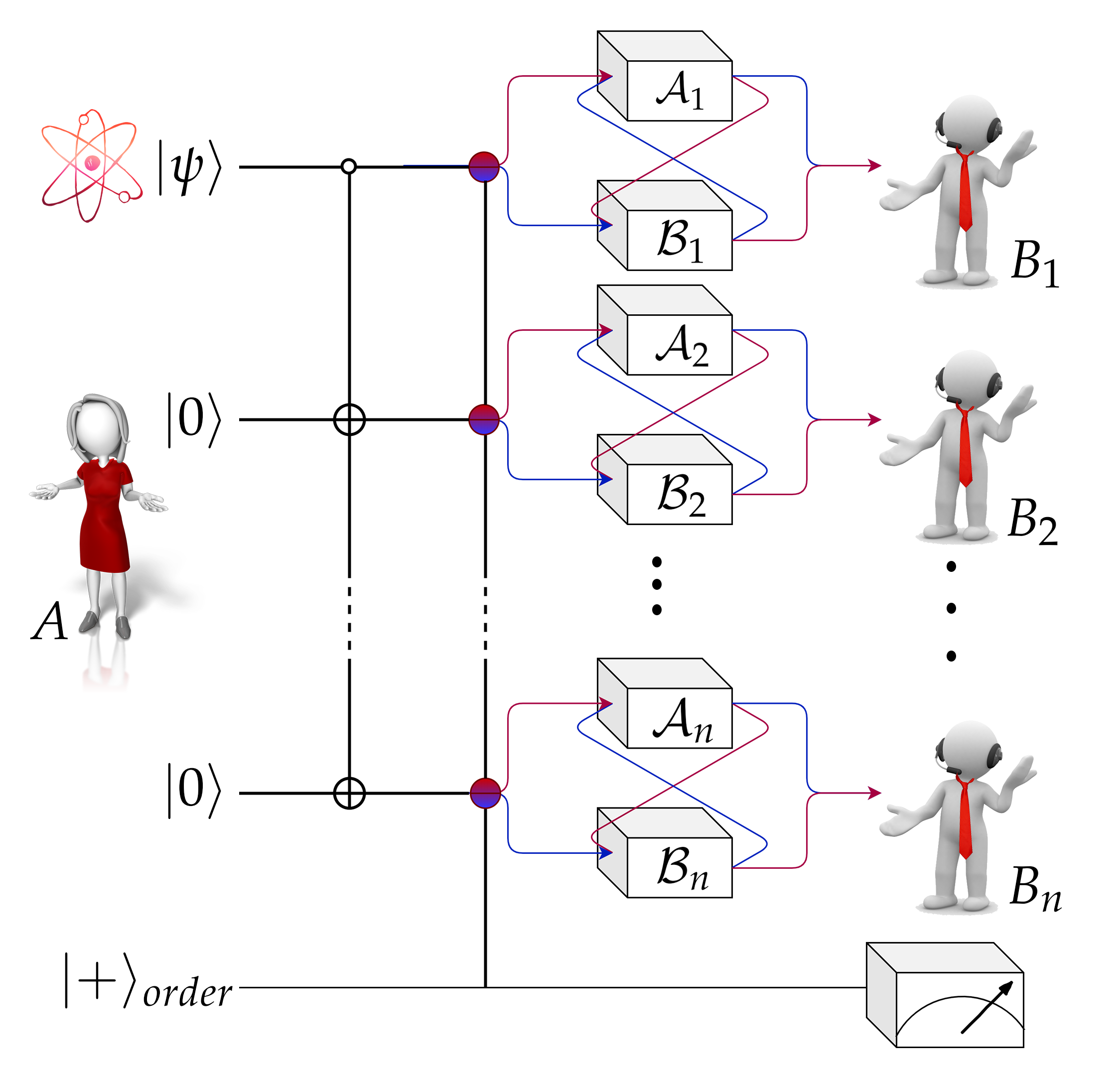}
		\caption{}
		\label{subfig:switch}
	\end{subfigure}
    \vspace{1em}
	\begin{subfigure}{\linewidth}
		\includegraphics[scale=0.4]{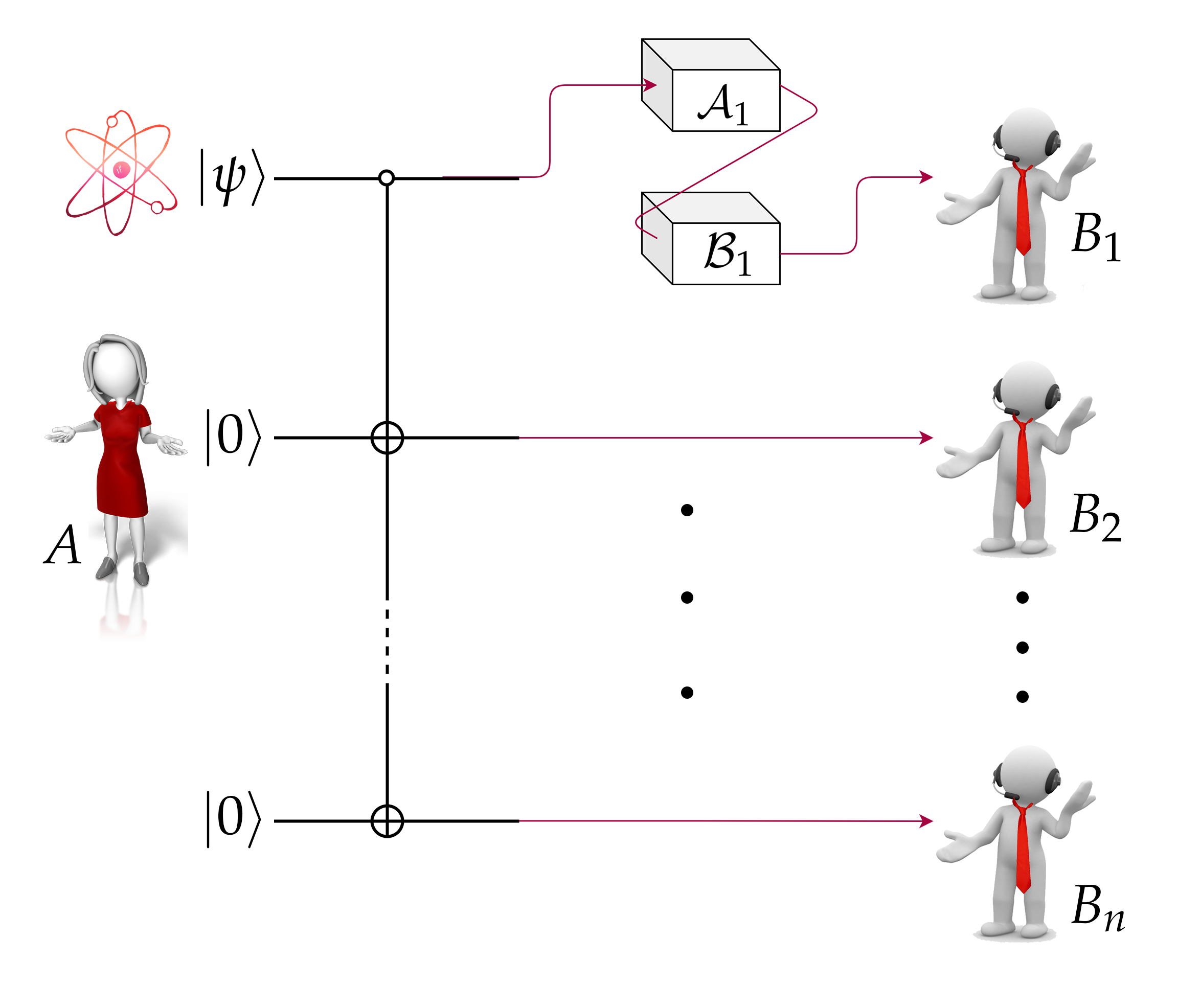}
		\caption{}
		\label{subfig:sidechannels}
	\end{subfigure}
		\caption{Random-receiver quantum communication task with indefinite and definite ordering of noisy channels. (a) Perfect protocol with indefinite order: Alice encodes the unknown qubit state $\psi$ in a $n$-partite GHZ state and sends through the noisy channels. The order qubit controls the order in which the subsystems pass through noisy channels, i.e. either $\{{\map A}_i\}$ before $\{{\map B}_i\}$ or $\{{\map B}_i\}$ before $\{{\map A}_i\}$. The order qubit has been prepared in $|+\rangle$ state, which is a superposition of these two orderings. Finally, the order qubit is measured and depending on the outcome the spatially separated Bobs apply local operations to establish GHZ correlation among themselves. Now even with LOCC the unknown state can be perfectly retrieved at any of the randomly chosen Bob's lab. (b) Even if Alice shares $n-1$ noiseless channels with $(n-1)$ Bobs and the noisy channels ${\map A}_1$, ${\map B}_1$ in a fixed order with the other Bob, the task cannot be perfectly accomplished.}
\end{figure}

In contrast with the above observations, in the following we will show that random-receiver quantum communication {\em can} take place when multiple entanglement-breaking channels are applied in a superposition of alternative orders.  Suppose that the quantum communication between the sender and the $i$-th receiver takes place through two channels $\map A_i$ and $\map B_i$, and that  the order of application of the two channels is entangled with the state of a control qubit, which we call the {\em order qubit}. In this scenario, illustrated in Figure \ref{subfig:switch}, we show that perfect random-receiver quantum communication is possible even if all the channels $(\map A_i ,\map B_i)_{i=1}^n$ are entanglement-breaking, provided that the order qubit is accessible through measurements, and that the outcome of a binary measurement  is sent to the chosen receiver.    In other words, the indefiniteness of the order enables $n$-party random-receiver quantum communication  using only entanglement-breaking channels  and one bit of classical communication  to one of the parties. 

Remarkably, this phenomenon cannot be reproduced in a scenario where the the order of the channels is definite and the sender can send  quantum data  to one of the parties, as illustrated in Figure \ref{subfig:sidechannels}.  In other words, the  access  to the qubit that determines the order is  a more powerful resource than the noiseless transmission of quantum data from the sender to one of the parties.   In fact, we prove an even stronger result: classical communication of the measurement outcomes on the order qubit is a more powerful resource than noiseless quantum communication to  $n-1$ parties.  
Achieving random-receiver quantum communication in the scenario of Figure \ref{subfig:sidechannels} requires at least one qubit of noiseless quantum communication to each of the $n$ parties.  
Our results show that the order qubit can unlock quantum communication to a randomly chosen receiver. The unlocking takes place thanks to the correlations between the order qubits and the output of the noisy channels connecting the sender to the receivers.   In contrast, any  noiseless quantum communication channel from the sender to a given receiver does not establish correlations with the output of the other receivers.  As a consequence, the only way to achieve random-receiver quantum communication through the addition of noiseless communication is to have one noiseless communication channel for each of the $n$ receivers.

{\em Conditions for random-access quantum communication.}    We first show that the noiseless protocol provided in the introduction is optimal among one-way protocols, that is, protocols consisting of a single round of classical communication to the chosen receiver.   
\begin{theo}\label{theo:oneway}
Every one-way protocol for random-receiver  communication of a $d$-dimensional quantum message requires each of the channels $(\map C_i)_{i=1}^n$ to have a quantum capacity of at least $\log d$ qubits. 
\end{theo} 
 The proof is provided in  Appendix \ref{appena}.  In particular, Theorem \ref{theo:oneway} implies that random-receiver quantum communication cannot take place when some  of the channels $(\map C_i)_{i=1}^n$ are entanglement-breaking. We recall that  entanglement-breaking channels are of the measure-and-prepare form $\map C  (\rho)   = \sum_j    \Tr [  M_j  \rho]    \,   \rho_j$, where $(M_j)$ is a quantum measurement and $\{\rho_j\}$ is a set of output states \cite{Holevo98}.  Entanglement-breaking channels are the prototype of channels with zero quantum capacity, and therefore they cannot achieve random-receiver quantum communication.    

  In the rest of the paper, we will focus on the scenario where {\em all}  channels are entanglement-breaking, and ask which additional resources should be added in order to enable random-receiver quantum communication.  In the basic model of Figure \ref{fig:basic}, we replace each entangelment-breaking channel $\map C_i$ with a new channel $\map C_i\otimes \map S_i$, where $\map S_i$ is an additional channel from the sender to the $i$-th receiver. For simplicity, we assume that each side-channel $\map S_i$ acts on  a quantum system of dimension $d$, equal to the dimension of the quantum message.  In this setting, we prove that random-receiver communication is possible if and only if each side-channel is noiseless. 
  
 \begin{theo}\label{theo:locc}
 Random-receiver quantum communication with entanglement-breaking channels $(\map C_i)_{i=1}^n$ and side-channels $(\map S_i)_{i=1}^n$ is possible if and only if all  side-channels are  noiseless. 
 \end{theo} 
 The proof is provided in Appendix \ref{appenb}.   In particular, Theorem \ref{theo:locc} shows that random-receiver quantum communication with entanglement-breaking channels is impossible even if one provides noiseless side-channels to $k<n$ receivers.  
   
  {\it Random-receiver quantum communication through the quantum SWITCH.}   Let ${\map A}:=\bigotimes_{i=1}^n{\map A}_i$ and ${\map B}:=\bigotimes_{i=1}^n{\map B}_i$  be two quantum channels, describing the noise experienced by the data transmitted by a sender to $n$ receivers. The  action  of the channels $\map A$ and $\map B$ in a superposition of two alternative orders is described by the quantum SWITCH, a higher-order map that transforms the pair of channels $(\map A,\map B)$ into a new quantum channels $\map S (\map A,  \map B)$, involving a control qubit that determines the order of application of channels $\map A$ and $\map B$. In its simples version, the quantum SWITCH produces the channel $\map S (\map A,  \map B)$ with Kraus operators 
  \begin{align}
S_{j k}   :  =  A_j B_k \otimes |0\>\<0|  +  B_k A_j \otimes |1\>\<1|  \, ,     
  \end{align}
  where $\{A_j\}$ and $\{B_k\}$ are Kraus representations for channels $\map A$ and $\map B$, respectively. It is easy to verify that the definition of channel $\map S (\map A, \map B)$ is independent of the choice of Kraus representations. 
    When the order qubit is initialized in the state $\omega$ we use the shorthand $\map S_\omega (\map A,  \map B)  (\rho):  =  \map S (\map A, \map B)     (  \rho \otimes \omega)$, and we call $\map S_\omega  (\map A,\map B)$ the {\em switched channel}. 
    
  When $\map A$ and $\map B$ are products of Pauli channels, the switched channel has the simple expression 
  \begin{align}\label{Cpm}
  \map S_\omega  (\map A,\map B)   =  p_+  \,  \map C_+  \,  \otimes \omega_+ +  p_- \, \map C_- \otimes \omega_+ \, ,
  \end{align}
  where $(p_+, p_-)$ are two probabilities, $\omega_+  := \omega$ and $\omega_-: = Z\omega Z$ are states of the order qubit, and $(\map C_+, \map C_-)$ are two quantum channels (see Appendix \ref{appenc} for the explicit expression).   In the following, we will focus on the case where all channels $(\map A_i)_{i=1}^n$ and $(\map B_i)_{i=1}^n$ are equal to the Pauli channel $\map N_{XY}$, defined by $\map N_{XY}  (\rho)   =  1/2  (  X \rho X  +  Y \rho Y)$.  This channel is entanglement-breaking and therefore cannot directly transmit quantum information.   However, we will see that the use of this channel in a superposition of orders achieves perfect quantum communication to a randomly chosen receiver. 
  
For simplicity, we illustrate the idea for $n=2$.  First, the sender encodes the message $|\psi\> = \alpha |0\>  + \beta|1\>$ into the state $|\psi_2\>:  =  \alpha |0\>|0\>  +  \beta |1\>|1\>$, as in the noiseless protocol.  Then, the sender sends the two qubits to receivers 1 and 2, using the channels $\map A=   \map N_{XY} \otimes \map N_{XY}$ and $\map B=   \map N_{XY} \otimes \map N_{XY}$  in a superposition of orders.  When  the order qubit is initialized in the state $|+\>  =  (|0\> + |1\> )/\sqrt 2$, 
the channels   $\map C_\pm$ in Eq. (\ref{Cpm}) are  
\begin{align}
\nonumber \map C_+  (\rho)    &=  \frac{\rho  +  (Z\otimes Z)  \rho  (Z\otimes  Z)} 2   \\
\map C_-   (\rho)    &  =  \frac { (  I \otimes Z) \rho  (I\otimes Z)  +  (Z\otimes I)  \rho  (Z\otimes I)}2  \,, 
\label{NS3}
	\end{align}
	and the probabilities $p_{\pm}$ are both equal to $1/2$.    The output states of the order qubit are either $\omega_+  =  |+\>\<+|$ or $\omega_-  =   |-\>\<-|$, with $|-\>  : =  (|0\>  -  |1\>)/\sqrt 2$.  Since these two states are orthogonal, a measurement on the order qubit   postselect one of the two channels $\map C_+$ or $\map C_-$.   Moreover, the channels $\map C_+$ and $\map C_-$ are equivalent under local unitary operations: for example, party 1 can turn channel $\map C_-$ into channel $\map C_+$ by applying the Pauli gate $Z$ on its qubit. If the outcome of the measurement on the order qubit is shared to the two receivers, they can ensure that their qubits have gone through the channel $\map C_+$.  Now, the  pure state $\rho=  |\psi_2\>\<\psi_2|$ is invariant under the action of channel $\map C_+$, and therefore it reaches the two receivers without any error. Hence, the two receivers end up with two qubits in the same state as in the noiseless protocol, and can achieve random-receiver quantum communication.  
	Summarizing, classical communication of the outcome of a measurement on the order qubit  enables  perfect random-receiver quantum communication.    

The above protocol can be generalized from $n=2$ to arbitrary numbers of receivers, as shown in Appendix \ref{append}.  The crucial feature of the protocol is that  access to a single qubit (the order qubit) is enough to unlock quantum communication to $n$ independent receivers. This feature cannot be reproduced by adding a qubit side-channel  in a causally ordered scenario.     In fact, Theorem \ref{theo:locc} implies that random-receiver quantum communication is impossible even if one adds any number $k<n$ of  qubit side-channels.  In short, the mere access to the order qubit is a more powerful resource than the access to $(n-1)$ qubit side-channels.

{\em Discussion.} Quantum communication with the assistance of the quantum SWITCH is similar to quantum communication with classical assistance from the environment \cite{Greg03,Greg04,Hayden05,Smolin05}.  In both cases, the access to a measurement outcome unlocks some quantum information that would be  inaccessible otherwise.  The analogy goes even further, because the quantum SWITCH of two Pauli channels $\map A $ and $ \map B$ is an extension of the quantum channel $\map A \map B$, that is, the channel that arises when channels $\map A$ and $\map B$ are  applied  in cascade in  a definite causal order.    Precisely, the channel $\map A \map B$   can be obtained from the switched channel $\map S_\omega (\map A, \map B)$ by discarding the order system.   From this point of view, the order qubit is indeed part of the environment of the channel $\map A \map B$,  and quantum communication with the assistance of the SWITCH is a special case of quantum communication  with classical assistance from the environment. The key difference is that, in the case of the quantum SWITCH, only a small part of the environment needs to be accessible, while in the other examples of quantum communication with the assistance of   environment it is generally assumed that the whole environment be accessible.  

Another class of communication protocols that exhibit similarities with  the quantum SWITCH are the communication protocols using controlled operations  before and after the communication channels  \cite{Guerin18}.   Like the quantum SWITCH, these protocols  use a control qubit, which determines the choice of operations performed on the input and output of  the communication channels.  The key difference with the quantum SWITCH is that such protocols generally transfer information to the control system in a way that bypasses the original channels  \cite{Chiribella18,Kristjansson19}.   In contrast, in all the protocols considered  in the literature, the quantum SWITCH does not deposit information into the order qubit. 
  For protocols involving Pauli channels, this feature is evident from Eq. (\ref{Cpm}), where the states $\omega_\pm$ of the order qubit are independent of the message, and so are the probabilities $p_\pm$ (see Appendix \ref{appenc} for the explicit expression).   

We observe that, if we allow arbitrary controlled operations before and after the noisy channels,  then protocols for random-receiver quantum communication with entanglement-breaking channels can be constructed also in the causally ordered scenario.  
This is because controlled operations can be used    {\em (i)} to transfer information directly from the message to the control qubit, bypassing the noisy channels $\map A$ and $\map B$, and {\em (ii)}  to generate the generalized GHZ state $\alpha |0\>^{\otimes n}   + \beta  |1\>^{\otimes n}$ from the state of the control qubit,  evading  the locality restriction  that affects the receivers.  An example of protocol that achieves random-receiver communication through controlled operations in a definite causal order is presented in Appendix \ref{appene}. 

The possibility of random-receiver quantum communication through controlled operations in a definite order can be interpreted in two ways. On the one hand, controlled operations can generate entanglement among the $n$ receivers, and therefore appear to be too powerful to be interesting in the problem of random-receiver quantum communication, where locality in space is an essential feature of the problem. On the other hand, controlled operations have some similarity with the quantum SWITCH, which can be regarded as a controlled {\tt SWAP} operation {\em in time}.  Controlled {\tt SWAP}  operations are a special subset of the set of all controlled operations, and one may wonder whether this special subset can reproduce the features of the quantum SWITCH.  Interestingly, the answer is negative:  in Appendix \ref{appene} we show that no controlled routing of the inputs and outputs of channels $\map A  =  \map B  = \map N_{XY}^{\otimes n}$ permits random-receiver quantum communication for odd $n$. 

{\em Photonic simulation of the random-receiver quantum communication task.} Quantum-SWITCH has recently been simulated in several photonic setups \cite{Procopio15,Rubino17,Goswami18(1),Guo20}. For instance, in the scheme of Ref.\cite{Goswami18(1)} photon's transverse spatial mode behaves as the target system evolving under two quantum operations whose relative order is controlled by photon's polarization degrees of freedom (DOF). For implementing random-receiver quantum communication through quantum SWITCH in photonic setup we require more than two DOFs to be considered at a time with one of them playing the role of order system. In the present context we assume that the sender possesses advanced optical devices that allow her to apply any joint (entangled) quantum operation on multiple DOFs of the photon, whereas the receiver can address each DOF individually. This assumption effectively mimics the scenario of random-receiver quantum communication with different DOFs playing the role of different spatially separated Bobs. In which DOF the quantum information has to be reproduced is decided at a later time after the DOFs evolve through noisy processes. Multiple DOFs of photon, such as polarization, spatial-mode, orbit-angular-momentum, time-bin and frequency have already been addressed simultaneously in different photonic experiments \cite{Parigi15,Wang15,Deng17,Luo16}. The proposed random-receiver quantum communication task thus welcomes an inquisitive conglomeration of presently available quantum optical devices to demonstrate a novel information theoretic advantage of indefinite causal order. 
 
{\em Conclusions.} Coherent control of orders/paths of quantum process has gained much of recent interests as it finds useful applications in quantum communication tasks \cite{Ebler18,Abbott18,Kar18}. 
To what extent these advantages are specific to superpositions of causal orders, rather than being generic to
other forms of coherent superpositions of communication protocols, is currently a matter of debate \cite{Chiribella18,Guerin18,Kristjansson19}. In particular, the advantage of coherent control of orders in time over that of paths in space is achieved under the distinct role of external and internal degrees of freedoms in communication task. In this regard the present work is quite important.
Here we have introduced a novel generalization of quantum communication task and established advantage of indefinite ordering of quantum processes over coherently controlled processes with fixed order. Importantly, this advantage implies that access to a qubit system, controlling the order of quantum processes, is a more powerful resource than $(n-1)$ qubit side-channels for any natural number $n>1$. Present study also opens up potential use of indefinite causal order in distributed protocols, such as multipartite quantum state transfer, quantum network, and entanglement distribution \cite{Briegel08,Cirac97} which have enormous practical relevance in the emerging new technology of quantum internet \cite{Kimble08,Wehner18}. 

\begin{acknowledgments}
The authors acknowledge discussion with G. Kar. AGM would like to thank Fabio Costa, Philippe Guérin, Cyril Branciard and \v Caslav Brukner for fruitful discussion during the poster session at "Causality in quantum world: harnessing quantum effects in causal inference problems" held at Anacapri, Italy.
This work is supported by the National Natural Science Foundation of China through grant~11675136, the Croucher Foundation, the Canadian Institute for Advanced Research~(CIFAR), the Hong Research Grant Council through grant~17307719 and the ID 61466 grant from the John Templeton Foundation, as part of the “The Quantum Information Structure of Spacetime (QISS)" Project (qiss.fr). The opinions expressed in this publication are those of the author and do not necessarily reflect the views of the John Templeton Foundation. MB acknowledges research grant through INSPIRE-faculty fellowship from the Department of Science and Technology, Government of India.
\end{acknowledgments}  

\bibliography{rrqc}

\begin{thebibliography}{25}%
\makeatletter
\providecommand \@ifxundefined [1]{%
 \@ifx{#1\undefined}
}%
\providecommand \@ifnum [1]{%
 \ifnum #1\expandafter \@firstoftwo
 \else \expandafter \@secondoftwo
 \fi
}%
\providecommand \@ifx [1]{%
 \ifx #1\expandafter \@firstoftwo
 \else \expandafter \@secondoftwo
 \fi
}%
\providecommand \natexlab [1]{#1}%
\providecommand \enquote  [1]{``#1''}%
\providecommand \bibnamefont  [1]{#1}%
\providecommand \bibfnamefont [1]{#1}%
\providecommand \citenamefont [1]{#1}%
\providecommand \href@noop [0]{\@secondoftwo}%
\providecommand \href [0]{\begingroup \@sanitize@url \@href}%
\providecommand \@href[1]{\@@startlink{#1}\@@href}%
\providecommand \@@href[1]{\endgroup#1\@@endlink}%
\providecommand \@sanitize@url [0]{\catcode `\\12\catcode `\$12\catcode
  `\&12\catcode `\#12\catcode `\^12\catcode `\_12\catcode `\%12\relax}%
\providecommand \@@startlink[1]{}%
\providecommand \@@endlink[0]{}%
\providecommand \url  [0]{\begingroup\@sanitize@url \@url }%
\providecommand \@url [1]{\endgroup\@href {#1}{\urlprefix }}%
\providecommand \urlprefix  [0]{URL }%
\providecommand \Eprint [0]{\href }%
\providecommand \doibase [0]{http://dx.doi.org/}%
\providecommand \selectlanguage [0]{\@gobble}%
\providecommand \bibinfo  [0]{\@secondoftwo}%
\providecommand \bibfield  [0]{\@secondoftwo}%
\providecommand \translation [1]{[#1]}%
\providecommand \BibitemOpen [0]{}%
\providecommand \bibitemStop [0]{}%
\providecommand \bibitemNoStop [0]{.\EOS\space}%
\providecommand \EOS [0]{\spacefactor3000\relax}%
\providecommand \BibitemShut  [1]{\csname bibitem#1\endcsname}%
\let\auto@bib@innerbib\@empty
\bibitem [{\citenamefont {Kent}(2013)}]{Kent}%
  \BibitemOpen
  \bibfield  {author} {\bibinfo {author} {\bibfnamefont {Adrian}\ \bibnamefont
  {Kent}},\ }\bibfield  {title} {\enquote {\bibinfo {title} {A no-summoning
  theorem in relativistic quantum theory},}\ }\href
  {https://link.springer.com/article/10.1007/s11128-012-0431-6} {\bibfield
  {journal} {\bibinfo  {journal} {Quantum Information Processing}\ }\textbf
  {\bibinfo {volume} {12}},\ \bibinfo {pages} {1023–1032} (\bibinfo {year}
  {2013})}\BibitemShut {NoStop}%
\bibitem [{\citenamefont {Kent}(2012)}]{Kent12}%
  \BibitemOpen
  \bibfield  {author} {\bibinfo {author} {\bibfnamefont {Adrian}\ \bibnamefont
  {Kent}},\ }\bibfield  {title} {\enquote {\bibinfo {title} {Quantum tasks in
  {M}inkowski space},}\ }\href {\doibase 10.1088/0264-9381/29/22/224013}
  {\bibfield  {journal} {\bibinfo  {journal} {Classical and Quantum Gravity}\
  }\textbf {\bibinfo {volume} {29}},\ \bibinfo {pages} {224013} (\bibinfo
  {year} {2012})}\BibitemShut {NoStop}%
\bibitem [{\citenamefont {Holevo}(1998)}]{Holevo98}%
  \BibitemOpen
  \bibfield  {author} {\bibinfo {author} {\bibfnamefont {A.~S.}\ \bibnamefont
  {Holevo}},\ }\bibfield  {title} {\enquote {\bibinfo {title} {Quantum coding
  theorems},}\ }\href {http://dx.doi.org/10.1070/RM1998v053n06ABEH000091}
  {\bibfield  {journal} {\bibinfo  {journal} {Russian Mathematical Surveys}\
  }\textbf {\bibinfo {volume} {53}},\ \bibinfo {pages} {1295} (\bibinfo {year}
  {1998})}\BibitemShut {NoStop}%
\bibitem [{\citenamefont {Gregoratti}\ and\ \citenamefont
  {Werner}(2003)}]{Greg03}%
  \BibitemOpen
  \bibfield  {author} {\bibinfo {author} {\bibfnamefont {M.}~\bibnamefont
  {Gregoratti}}\ and\ \bibinfo {author} {\bibfnamefont {R.~F.}\ \bibnamefont
  {Werner}},\ }\bibfield  {title} {\enquote {\bibinfo {title} {Quantum lost and
  found},}\ }\href
  {https://www.tandfonline.com/doi/abs/10.1080/09500340308234541} {\bibfield
  {journal} {\bibinfo  {journal} {Journal of Modern Optics}\ }\textbf {\bibinfo
  {volume} {50}},\ \bibinfo {pages} {915--933} (\bibinfo {year}
  {2003})}\BibitemShut {NoStop}%
\bibitem [{\citenamefont {Gregoratti}\ and\ \citenamefont
  {Werner}(2004)}]{Greg04}%
  \BibitemOpen
  \bibfield  {author} {\bibinfo {author} {\bibfnamefont {M.}~\bibnamefont
  {Gregoratti}}\ and\ \bibinfo {author} {\bibfnamefont {R.~F.}\ \bibnamefont
  {Werner}},\ }\bibfield  {title} {\enquote {\bibinfo {title} {On quantum
  error-correction by classical feedback in discrete time},}\ }\href
  {https://doi.org/10.1063/1.1758320} {\bibfield  {journal} {\bibinfo
  {journal} {J. Math. Phys.}\ }\textbf {\bibinfo {volume} {45}},\ \bibinfo
  {pages} {2600} (\bibinfo {year} {2004})}\BibitemShut {NoStop}%
\bibitem [{\citenamefont {Hayden}\ and\ \citenamefont {King}(2005)}]{Hayden05}%
  \BibitemOpen
  \bibfield  {author} {\bibinfo {author} {\bibfnamefont {P.}~\bibnamefont
  {Hayden}}\ and\ \bibinfo {author} {\bibfnamefont {C.}~\bibnamefont {King}},\
  }\bibfield  {title} {\enquote {\bibinfo {title} {Correcting quantum channels
  by measuring the environment},}\ }\href {https://doi.org/10.26421/QIC5.2}
  {\bibfield  {journal} {\bibinfo  {journal} {Quan. Inf. Comp.}\ }\textbf
  {\bibinfo {volume} {5}},\ \bibinfo {pages} {156} (\bibinfo {year}
  {2005})}\BibitemShut {NoStop}%
\bibitem [{\citenamefont {Smolin}\ \emph {et~al.}(2005)\citenamefont {Smolin},
  \citenamefont {Verstraete},\ and\ \citenamefont {Winter}}]{Smolin05}%
  \BibitemOpen
  \bibfield  {author} {\bibinfo {author} {\bibfnamefont {J.~A.}\ \bibnamefont
  {Smolin}}, \bibinfo {author} {\bibfnamefont {F.}~\bibnamefont {Verstraete}},
  \ and\ \bibinfo {author} {\bibfnamefont {A.}~\bibnamefont {Winter}},\
  }\bibfield  {title} {\enquote {\bibinfo {title} {Entanglement of assistance
  and multipartite state distillation},}\ }\href
  {https://link.aps.org/doi/10.1103/PhysRevA.72.052317} {\bibfield  {journal}
  {\bibinfo  {journal} {Phys. Rev. A}\ }\textbf {\bibinfo {volume} {72}},\
  \bibinfo {pages} {052317} (\bibinfo {year} {2005})}\BibitemShut {NoStop}%
\bibitem [{\citenamefont {Guérin}\ \emph {et~al.}(2019)\citenamefont
  {Guérin}, \citenamefont {Rubino},\ and\ \citenamefont {Brukner}}]{Guerin18}%
  \BibitemOpen
  \bibfield  {author} {\bibinfo {author} {\bibfnamefont {A.}~\bibnamefont
  {Guérin}}, \bibinfo {author} {\bibfnamefont {G.}~\bibnamefont {Rubino}}, \
  and\ \bibinfo {author} {\bibfnamefont {Č}~\bibnamefont {Brukner}},\
  }\bibfield  {title} {\enquote {\bibinfo {title} {Communication through
  quantum-controlled noise},}\ }\href
  {https://doi.org/10.1103/PhysRevA.99.062317} {\bibfield  {journal} {\bibinfo
  {journal} {Phys. Rev. A}\ }\textbf {\bibinfo {volume} {99}},\ \bibinfo
  {pages} {062317} (\bibinfo {year} {2019})}\BibitemShut {NoStop}%
\bibitem [{\citenamefont {Chiribella}\ and\ \citenamefont
  {Kristjánsson}(2019)}]{Chiribella18}%
  \BibitemOpen
  \bibfield  {author} {\bibinfo {author} {\bibfnamefont {G.}~\bibnamefont
  {Chiribella}}\ and\ \bibinfo {author} {\bibfnamefont {H.}~\bibnamefont
  {Kristjánsson}},\ }\bibfield  {title} {\enquote {\bibinfo {title} {Quantum
  shannon theory with superpositions of trajectories},}\ }\href
  {https://doi.org/10.6084/m9.figshare.c.4470896.v1} {\bibfield  {journal}
  {\bibinfo  {journal} {Proceedings of the Royal Society A}\ }\textbf {\bibinfo
  {volume} {475}},\ \bibinfo {pages} {20180903} (\bibinfo {year}
  {2019})}\BibitemShut {NoStop}%
\bibitem [{\citenamefont {Kristjánsson}\ \emph {et~al.}()\citenamefont
  {Kristjánsson}, \citenamefont {Salek}, \citenamefont {Ebler},\ and\
  \citenamefont {Chiribella}}]{Kristjansson19}%
  \BibitemOpen
  \bibfield  {author} {\bibinfo {author} {\bibfnamefont {H.}~\bibnamefont
  {Kristjánsson}}, \bibinfo {author} {\bibfnamefont {S.}~\bibnamefont
  {Salek}}, \bibinfo {author} {\bibfnamefont {D.}~\bibnamefont {Ebler}}, \ and\
  \bibinfo {author} {\bibfnamefont {G.}~\bibnamefont {Chiribella}},\ }\bibfield
   {title} {\enquote {\bibinfo {title} {Resource theories of communication with
  quantum superpositions of processes},}\ }\href
  {https://arxiv.org/abs/1910.08197} {\ }\Eprint
  {http://arxiv.org/abs/1910.08197} {arXiv:1910.08197} \BibitemShut {NoStop}%
\bibitem [{\citenamefont {et~al.}(2015{\natexlab{a}})}]{Procopio15}%
  \BibitemOpen
  \bibfield  {author} {\bibinfo {author} {\bibfnamefont {L.~M.~Procopio}\
  \bibnamefont {et~al.}},\ }\bibfield  {title} {\enquote {\bibinfo {title}
  {Experimental superposition of orders of quantum gates},}\ }\href
  {https://www.nature.com/articles/ncomms8913} {\bibfield  {journal} {\bibinfo
  {journal} {Nat. Commun.}\ }\textbf {\bibinfo {volume} {6}} (\bibinfo {year}
  {2015}{\natexlab{a}})}\BibitemShut {NoStop}%
\bibitem [{\citenamefont {al}(2017)}]{Rubino17}%
  \BibitemOpen
  \bibfield  {author} {\bibinfo {author} {\bibfnamefont {G.~Rubino}\
  \bibnamefont {al}},\ }\bibfield  {title} {\enquote {\bibinfo {title}
  {Experimental verification of an indefinite causal order},}\ }\href
  {http://advances.sciencemag.org/content/3/3/e1602589} {\bibfield  {journal}
  {\bibinfo  {journal} {Science Advances}\ }\textbf {\bibinfo {volume} {3}},\
  \bibinfo {pages} {e1602589} (\bibinfo {year} {2017})}\BibitemShut {NoStop}%
\bibitem [{\citenamefont {Goswami}\ \emph {et~al.}(2018)\citenamefont
  {Goswami}, \citenamefont {Giarmatzi}, \citenamefont {Kewming}, \citenamefont
  {Costa}, \citenamefont {Branciard}, \citenamefont {Romero},\ and\
  \citenamefont {White}}]{Goswami18(1)}%
  \BibitemOpen
  \bibfield  {author} {\bibinfo {author} {\bibfnamefont {K.}~\bibnamefont
  {Goswami}}, \bibinfo {author} {\bibfnamefont {C.}~\bibnamefont {Giarmatzi}},
  \bibinfo {author} {\bibfnamefont {M.}~\bibnamefont {Kewming}}, \bibinfo
  {author} {\bibfnamefont {F.}~\bibnamefont {Costa}}, \bibinfo {author}
  {\bibfnamefont {C.}~\bibnamefont {Branciard}}, \bibinfo {author}
  {\bibfnamefont {J.}~\bibnamefont {Romero}}, \ and\ \bibinfo {author}
  {\bibfnamefont {A.~G.}\ \bibnamefont {White}},\ }\bibfield  {title} {\enquote
  {\bibinfo {title} {Indefinite causal order in a quantum switch},}\ }\href
  {https://doi.org/10.1103/PhysRevLett.121.090503} {\bibfield  {journal}
  {\bibinfo  {journal} {Phys. Rev. Lett.}\ }\textbf {\bibinfo {volume} {121}},\
  \bibinfo {pages} {090503} (\bibinfo {year} {2018})}\BibitemShut {NoStop}%
\bibitem [{\citenamefont {Guo}\ \emph {et~al.}(2020)\citenamefont {Guo},
  \citenamefont {Hu}, \citenamefont {Hou}, \citenamefont {Cao}, \citenamefont
  {Cui}, \citenamefont {Liu}, \citenamefont {Huang}, \citenamefont {Li},
  \citenamefont {Guo},\ and\ \citenamefont {Chiribella}}]{Guo20}%
  \BibitemOpen
  \bibfield  {author} {\bibinfo {author} {\bibfnamefont {Y.}~\bibnamefont
  {Guo}}, \bibinfo {author} {\bibfnamefont {X.-M.}\ \bibnamefont {Hu}},
  \bibinfo {author} {\bibfnamefont {Z.-B.}\ \bibnamefont {Hou}}, \bibinfo
  {author} {\bibfnamefont {H.}~\bibnamefont {Cao}}, \bibinfo {author}
  {\bibfnamefont {J.-M.}\ \bibnamefont {Cui}}, \bibinfo {author} {\bibfnamefont
  {B.-H.}\ \bibnamefont {Liu}}, \bibinfo {author} {\bibfnamefont {Y.-F.}\
  \bibnamefont {Huang}}, \bibinfo {author} {\bibfnamefont {C.-F.}\ \bibnamefont
  {Li}}, \bibinfo {author} {\bibfnamefont {G.-C.}\ \bibnamefont {Guo}}, \ and\
  \bibinfo {author} {\bibfnamefont {G.}~\bibnamefont {Chiribella}},\ }\bibfield
   {title} {\enquote {\bibinfo {title} {Experimental transmission of quantum
  information using a superposition of causal orders},}\ }\href
  {https://doi.org/10.1103/PhysRevLett.124.030502} {\bibfield  {journal}
  {\bibinfo  {journal} {Phys. Rev. Lett.}\ }\textbf {\bibinfo {volume} {124}},\
  \bibinfo {pages} {030502} (\bibinfo {year} {2020})}\BibitemShut {NoStop}%
\bibitem [{\citenamefont {et~al.}(2015{\natexlab{b}})}]{Parigi15}%
  \BibitemOpen
  \bibfield  {author} {\bibinfo {author} {\bibfnamefont {V.~Parigi}\
  \bibnamefont {et~al.}},\ }\bibfield  {title} {\enquote {\bibinfo {title}
  {Storage and retrieval of vector beams of light in a
  multiple-degree-of-freedom quantum memory},}\ }\href
  {https://www.nature.com/articles/ncomms8706} {\bibfield  {journal} {\bibinfo
  {journal} {Nat. Commun.}\ }\textbf {\bibinfo {volume} {6}},\ \bibinfo {pages}
  {7706} (\bibinfo {year} {2015}{\natexlab{b}})}\BibitemShut {NoStop}%
\bibitem [{\citenamefont {et~al}(2015)}]{Wang15}%
  \BibitemOpen
  \bibfield  {author} {\bibinfo {author} {\bibfnamefont {Xi-Lin~Wang}\
  \bibnamefont {et~al}},\ }\bibfield  {title} {\enquote {\bibinfo {title}
  {Quantum teleportation of multiple degrees of freedom of a single photon},}\
  }\href {https://www.nature.com/articles/nature14246} {\bibfield  {journal}
  {\bibinfo  {journal} {Nature}\ }\textbf {\bibinfo {volume} {518}},\ \bibinfo
  {pages} {516} (\bibinfo {year} {2015})}\BibitemShut {NoStop}%
\bibitem [{\citenamefont {Deng}\ \emph {et~al.}(2017)\citenamefont {Deng},
  \citenamefont {Ren},\ and\ \citenamefont {Li}}]{Deng17}%
  \BibitemOpen
  \bibfield  {author} {\bibinfo {author} {\bibfnamefont {Fu-Guo}\ \bibnamefont
  {Deng}}, \bibinfo {author} {\bibfnamefont {Bao-Cang}\ \bibnamefont {Ren}}, \
  and\ \bibinfo {author} {\bibfnamefont {Xi-Han}\ \bibnamefont {Li}},\
  }\bibfield  {title} {\enquote {\bibinfo {title} {Quantum hyperentanglement
  and its applications in quantum information processing},}\ }\href
  {https://doi.org/10.1016/j.scib.2016.11.007} {\bibfield  {journal} {\bibinfo
  {journal} {Science Bulletin}\ }\textbf {\bibinfo {volume} {62}},\ \bibinfo
  {pages} {46} (\bibinfo {year} {2017})}\BibitemShut {NoStop}%
\bibitem [{\citenamefont {Luo}\ \emph {et~al.}(2016)\citenamefont {Luo},
  \citenamefont {Li}, \citenamefont {Lai},\ and\ \citenamefont {Wang}}]{Luo16}%
  \BibitemOpen
  \bibfield  {author} {\bibinfo {author} {\bibfnamefont {Ming-Xing}\
  \bibnamefont {Luo}}, \bibinfo {author} {\bibfnamefont {Hui-Ran}\ \bibnamefont
  {Li}}, \bibinfo {author} {\bibfnamefont {Hong}\ \bibnamefont {Lai}}, \ and\
  \bibinfo {author} {\bibfnamefont {Xiaojun}\ \bibnamefont {Wang}},\ }\bibfield
   {title} {\enquote {\bibinfo {title} {Quantum computation based on photons
  with three degrees of freedom},}\ }\href
  {https://www.nature.com/articles/srep25977} {\bibfield  {journal} {\bibinfo
  {journal} {Scientific Reports}\ }\textbf {\bibinfo {volume} {6}},\ \bibinfo
  {pages} {25977} (\bibinfo {year} {2016})}\BibitemShut {NoStop}%
\bibitem [{\citenamefont {Ebler}\ \emph {et~al.}(2018)\citenamefont {Ebler},
  \citenamefont {Salek},\ and\ \citenamefont {Chiribella}}]{Ebler18}%
  \BibitemOpen
  \bibfield  {author} {\bibinfo {author} {\bibfnamefont {D.}~\bibnamefont
  {Ebler}}, \bibinfo {author} {\bibfnamefont {S.}~\bibnamefont {Salek}}, \ and\
  \bibinfo {author} {\bibfnamefont {G.}~\bibnamefont {Chiribella}},\ }\bibfield
   {title} {\enquote {\bibinfo {title} {Enhanced communication with the
  assistance of indefinite causal order},}\ }\href
  {https://doi.org/10.1103/PhysRevLett.120.120502} {\bibfield  {journal}
  {\bibinfo  {journal} {Phys. Rev. Lett.}\ }\textbf {\bibinfo {volume} {120}},\
  \bibinfo {pages} {120502} (\bibinfo {year} {2018})}\BibitemShut {NoStop}%
\bibitem [{\citenamefont {Abbott}\ \emph {et~al.}()\citenamefont {Abbott},
  \citenamefont {Wechs}, \citenamefont {Horsman}, \citenamefont {Mhalla},\ and\
  \citenamefont {Branciard}}]{Abbott18}%
  \BibitemOpen
  \bibfield  {author} {\bibinfo {author} {\bibfnamefont {A.~A.}\ \bibnamefont
  {Abbott}}, \bibinfo {author} {\bibfnamefont {J.}~\bibnamefont {Wechs}},
  \bibinfo {author} {\bibfnamefont {D.}~\bibnamefont {Horsman}}, \bibinfo
  {author} {\bibfnamefont {M.}~\bibnamefont {Mhalla}}, \ and\ \bibinfo {author}
  {\bibfnamefont {C.}~\bibnamefont {Branciard}},\ }\bibfield  {title} {\enquote
  {\bibinfo {title} {Communication through coherent control of quantum
  channels},}\ }\href {https://arxiv.org/abs/1810.09826} {\ }\Eprint
  {http://arxiv.org/abs/1810.09826} {arXiv:1810.09826} \BibitemShut {NoStop}%
\bibitem [{\citenamefont {Chiribella}\ \emph {et~al.}()\citenamefont
  {Chiribella}, \citenamefont {Banik}, \citenamefont {Bhattacharya},
  \citenamefont {Guha}, \citenamefont {Alimuddin}, \citenamefont {Roy},
  \citenamefont {Saha}, \citenamefont {Agrawal},\ and\ \citenamefont
  {Kar}}]{Kar18}%
  \BibitemOpen
  \bibfield  {author} {\bibinfo {author} {\bibfnamefont {G.}~\bibnamefont
  {Chiribella}}, \bibinfo {author} {\bibfnamefont {M.}~\bibnamefont {Banik}},
  \bibinfo {author} {\bibfnamefont {S.~S.}\ \bibnamefont {Bhattacharya}},
  \bibinfo {author} {\bibfnamefont {T.}~\bibnamefont {Guha}}, \bibinfo {author}
  {\bibfnamefont {M.}~\bibnamefont {Alimuddin}}, \bibinfo {author}
  {\bibfnamefont {A.}~\bibnamefont {Roy}}, \bibinfo {author} {\bibfnamefont
  {S.}~\bibnamefont {Saha}}, \bibinfo {author} {\bibfnamefont {S.}~\bibnamefont
  {Agrawal}}, \ and\ \bibinfo {author} {\bibfnamefont {G.}~\bibnamefont
  {Kar}},\ }\bibfield  {title} {\enquote {\bibinfo {title} {Indefinite causal
  order enables perfect quantum communication with zero capacity channel},}\
  }\href {https://arxiv.org/abs/1810.10457} {\ }\Eprint
  {http://arxiv.org/abs/1810.10457} {arXiv:1810.10457} \BibitemShut {NoStop}%
\bibitem [{\citenamefont {Briegel}\ \emph {et~al.}(2000)\citenamefont
  {Briegel}, \citenamefont {van Enk}, \citenamefont {Cirac},\ and\
  \citenamefont {Zoller}}]{Briegel08}%
  \BibitemOpen
  \bibfield  {author} {\bibinfo {author} {\bibfnamefont {H.~J.}\ \bibnamefont
  {Briegel}}, \bibinfo {author} {\bibfnamefont {S.~J.}\ \bibnamefont {van
  Enk}}, \bibinfo {author} {\bibfnamefont {J.~I.}\ \bibnamefont {Cirac}}, \
  and\ \bibinfo {author} {\bibfnamefont {P.}~\bibnamefont {Zoller}},\
  }\href@noop {} {\emph {\bibinfo {title} {The Physics of Quantum
  Information}}},\ edited by\ \bibinfo {editor} {\bibfnamefont {Ekert A. \&
  Zeilinger~A.}\ \bibnamefont {Bouwmeester}, \bibfnamefont {D.}}\ (\bibinfo
  {publisher} {Springer},\ \bibinfo {address} {Berlin},\ \bibinfo {year}
  {2000})\BibitemShut {NoStop}%
\bibitem [{\citenamefont {Cirac}\ \emph {et~al.}(1997)\citenamefont {Cirac},
  \citenamefont {Zoller}, \citenamefont {Kimble},\ and\ \citenamefont
  {Mabuchi}}]{Cirac97}%
  \BibitemOpen
  \bibfield  {author} {\bibinfo {author} {\bibfnamefont {J.~I.}\ \bibnamefont
  {Cirac}}, \bibinfo {author} {\bibfnamefont {P.}~\bibnamefont {Zoller}},
  \bibinfo {author} {\bibfnamefont {H.~J.}\ \bibnamefont {Kimble}}, \ and\
  \bibinfo {author} {\bibfnamefont {H.}~\bibnamefont {Mabuchi}},\ }\bibfield
  {title} {\enquote {\bibinfo {title} {Quantum state transfer and entanglement
  distribution among distant nodes in a quantum network},}\ }\href
  {https://doi.org/10.1103/PhysRevLett.78.3221} {\bibfield  {journal} {\bibinfo
   {journal} {Phys. Rev. Lett.}\ }\textbf {\bibinfo {volume} {78}},\ \bibinfo
  {pages} {3221} (\bibinfo {year} {1997})}\BibitemShut {NoStop}%
\bibitem [{\citenamefont {Kimble}(2008)}]{Kimble08}%
  \BibitemOpen
  \bibfield  {author} {\bibinfo {author} {\bibfnamefont {H.~J.}\ \bibnamefont
  {Kimble}},\ }\bibfield  {title} {\enquote {\bibinfo {title} {The quantum
  internet},}\ }\href {https://doi.org/10.1038/nature07127} {\bibfield
  {journal} {\bibinfo  {journal} {Nature}\ }\textbf {\bibinfo {volume} {453}},\
  \bibinfo {pages} {1023} (\bibinfo {year} {2008})}\BibitemShut {NoStop}%
\bibitem [{\citenamefont {Wehner}\ \emph {et~al.}(2018)\citenamefont {Wehner},
  \citenamefont {Elkouss},\ and\ \citenamefont {Hanson}}]{Wehner18}%
  \BibitemOpen
  \bibfield  {author} {\bibinfo {author} {\bibfnamefont {S.}~\bibnamefont
  {Wehner}}, \bibinfo {author} {\bibfnamefont {D.}~\bibnamefont {Elkouss}}, \
  and\ \bibinfo {author} {\bibfnamefont {R.}~\bibnamefont {Hanson}},\
  }\bibfield  {title} {\enquote {\bibinfo {title} {Quantum internet: A vision
  for the road ahead},}\ }\href
  {http://science.sciencemag.org/content/362/6412/eaam9288} {\bibfield
  {journal} {\bibinfo  {journal} {Science}\ }\textbf {\bibinfo {volume}
  {362}},\ \bibinfo {pages} {9288} (\bibinfo {year} {2018})}\BibitemShut
  {NoStop}%
\end{thebibliography}%

\onecolumngrid
\appendix

\section{Proof of Theorem \ref{theo:oneway}} \label{appena} 

  The proof uses a general result, expressed in terms of the following definition: for a generic quantum channel $\map C_L$ of a generic quantum system $L$, we say that $\map C$ {\em can transmit a $d$-dimensional quantum system in a one-way protocol} if there exists an encoding channel $\map E_{LR}  :  L  (\spc H_S)  \to L( \spc H_L \otimes \spc H_R)$,  a measurement a measurement $(P_j)_{j}$ on system $R$, and a set of local operations $(\map D_j)_{j}$ on system $L$, such that  $\sum_{j}  \Tr_L[ (   \map D_j\map C_L \otimes P_j    )\map E_{LR}  (\rho)]   = \rho $ for every state $\rho  \in \St (\C^d)$.

\begin{prop}\label{prop:oneway}  
 If channel $\map C_L$ can transfer  the state of a $d$-dimensional quantum system  in a one-way protocol, then channel $\map C_L$ has quantum capacity of at least $\log d$. 
\end{prop}
 
 {\bf Proof.}  Defining $\map E_j:  =\Tr_L [  (\map I_R\otimes P_j )  \map E_{LR}] $, we obtain the equivalent condition $\sum_j    \map D_j  \map C_R  \map E_j  =  \map I_d$.    This condition is satisfied if and only if each term in the sum is proportional to the identity map, namely $\map D_j  \map C_R  \map E_j =  p_j \, \map I_S$ for some probability distribution $(p_j)_{j}$. Since $\map F_j$ and $\map C_L$ are trace-preserving, this condition implies that $\map E_j'  :  = \map E_j/p_j$ is trace-preserving.   Since the condition  $ \map D_j\map C_R\map E_j'  = \map I_d$ holds, the channel $\map C_R$ permits a perfect transmission of a $d$-dimensional system, and therefore has a quantum capacity of at least $\log d$ qubits.  \qed 
  
\medskip

{\bf Proof of Theorem \ref{theo:oneway}.}   Suppose that there exists a one-way protocol for  random-receiver quantum communication using channels $(\map C_i)_{i=1}^n$, and that the protocol can successfully transfer a $d$-dimensional quantum system to any of the receivers $(B_i)_{i=1}^n$.  Let $\map E:   L(\C^d)  \to  L (\spc H_1\otimes \cdots \otimes \spc H_n)$ be the encoding channel used in the protocol. For every $x\in \{1,\dots, n\}$, let  $(M^{(x)}_j)$ be the measurement performed by the $(n-1)$ parties other than party $x$, and let $(   \map B^{(x)}_j)$ be the conditional operations performed by party $x$.    We can then regard systems $B_x$ and $\bigotimes_{y\not = x}  B_y$  as systems $L$ and $R$ in Proposition \ref{prop:oneway}, with encoding channel $\map E_{LR} :=  (\map F_1 \otimes \cdots \otimes \map F_n)  \map E$ with $\map E_y  := \map C_y$ for $y\not =  x$ and $\map E_x  :  = \map I_x$.  Applying Proposition  \ref{prop:oneway}, we then obtain that channel $\map C_x$ must have a capacity of at least $\log d$ qubits. Since $x$ is an arbitrary number in $\{1,\dots, n\}$, every channel  in the set $(\map C_i)_{i=1}^n$ must have a capacity of at least $\log d$.  \qed

 \section{Proof of Theorem \ref{theo:locc}}  \label{appenb}

The proof uses a generalization of Proposition \ref{prop:oneway} to arbitrary separable protocols. 
For a generic quantum channel $\map S_L$ on a generic quantum system $L$, we say that $\map C$ {\em can transmit a $d$-dimensional quantum system in an separable protocol} if there exists a system $R$,  an encoding channel $\map E_{LR}  :  L  (\spc H_S)  \to L( \spc H_L \otimes \spc H_R)$,  and a separable channel $\map D   =  \sum_{j}   \map L_j  \otimes \map R_j$ where  $\map L_j:   L(  \spc H_L) \to L(\C^d)$ and $\map R_j  :  L (\spc H_R) \to \C$ are completely positive maps for every $j$, such that 
\begin{align}\label{condition}
\map D (\map S_L  \otimes \map  I_R) \map E    =  \map I_d\, ,   
\end{align}
where $\map I_R$ is the identity channel on system $R$.  

\begin{prop}\label{prop:sep}  
 If the input of channel $\map S_L$ is a $d$-dimensional quantum system and $\map S_L$ can transfer  the state of a $d$-dimensional quantum system   in  a separable protocol, then $\map S_L$ is a unitary channel. 
 \end{prop}

 {\bf Proof.}  Defining $\map E_j:  =  (\map I_L\otimes \map R_j )  \map E_{LR}] $, we can rewrite Eq. (\ref{condition})  as  $\sum_j    \map L_j  \map S_L  \map E_j  =  \map I_d$.    This condition is satisfied if and only if each term in the sum is proportional to the identity map, namely $\map D_j  \map C_R  \map E_j =  p_j \, \map I_S$ for some probability distribution $(p_j)_{j}$.  Using Kraus representations for the maps $\map L_j$, $\map S_L$ and $\map E_j$, we obtain the condition  $    F_{jk} S_{l} E_{jm}=  \lambda_{jklm} \,  I_d$ for suitable coefficients $\lambda_{jklm}$ satisfying the normalization condition $\sum_{j,k,l,n} |\lambda_{jkmn}|^2  =1$.  Due to the normalization condition, there must exist  values of the indices $(j,k,l,m)$  such that $\lambda_{jklm}\not =  0$.  For these values, the condition  $    F_{jk} S_{l} E_{jm}=  \lambda_{jklm} \,  I_d$ implies that the operator $F_{jk}$ is invertible, and that one has $S_{l}  E_{jm}  =  \lambda_{jkln} \,  F_{jk}^{-1}$.   Multiplying by $F_{jk}$ on both sides of the equation, we then obtain  
\begin{align}\label{correction}
S_{l}  \,  E_{jm}    F_{jk}   =  \lambda_{jklm} \, I_d \,.
\end{align} 
Now, define the completely positive map $\widetilde {\map E}$  by $ \widetilde {\map E}  (\rho):=  \sum_{j,k,m}   E_{jm}    F_{jk}   \rho  F_{jk}^\dag   E_{jm}^\dag $. Eq. (\ref{correction}) implies the relation $\map S_L\widetilde {\map E}  =  \map I_d$.   Since $\map S_L$ and $\map I_d$ are trace-preserving, also $\widetilde {\map E}$ must be trace-preserving. Hence, $\widetilde {\map E}$ is an invertible quantum channel and $\map S_L$ is its inverse. Since the input and output systems of these two channels have the same dimension, this means that both channels must be unitary. \qed 
 
 \medskip 
  
{\bf Proof of Theorem \ref{theo:locc}.}     Suppose that there exists a general LOCC  protocol for  random-receiver quantum communication using channels $(\map C_i)_{i=1}^n$ and  side-channels  $(\map S_i)_{i=1}^n$ acting on  $d$-dimensional quantum systems.    Let $\map E:   L(\C^d)  \to  L (\spc H_1'\otimes \cdots \otimes \spc H_n')$ be the encoding channel used in the protocol, with $\spc H_i'   :  = \spc H_{B_i}\otimes \spc H_{S_i}$, where $\spc H_i$ is the output of channel $  \map C_i$ and $S_i  = \C^d$ is the output of the side-channel $\map S_i$. 

 For every $x\in \{1,\dots, n\}$, we can regard systems $S_x$ and $B_x \otimes \left[\bigotimes_{y\not = x}  (B_y\otimes S_y)\right]$  as systems $L$ and $R$ in Proposition \ref{prop:sep}, with channel $\map S_L:=  \map S_x$ and encoding channel $\map E_{LR} :=  (\map F_1 \otimes \cdots \otimes \map F_n)  \map E$ with $\map E_y  := \map C_y \otimes \map S_j$ for $y\not =  x$ and $\map E_x  :  = \map C_{B_x} \otimes \map I_{S_x}$.  The original LOCC protocol can be regarded as a special case of separable protocol with respect to the bipartition $(L,R)$.  Applying Proposition  \ref{prop:oneway}, we then obtain that channel $\map S_x$ must be unitary.  Since $x$ is an arbitrary element of $\{1,\dots, n\}$, every channel  in the set $(\map S_i)_{i=1}^n$ must be unitary.   \qed

\section{Switching products of Pauli channels }\label{appenc}
Let ${\map E}:=\bigotimes_{k=1}^2{\map E}_k$ be the product of two Pauli channels,  given by $\mathcal{E}_1\equiv\{\sqrt{p_0}\mathbb{I},\sqrt{p_1}X,\sqrt{p_2}Y,\sqrt{p_3}Z\}$ and $\mathcal{E}_2\equiv\{\sqrt{q_0}\mathbb{I},\sqrt{q_1}X,\sqrt{q_2}Y,\sqrt{q_3}Z\}$, $\sum_{l=0}^3p_l^2=\sum_{l=0}^3q_l^2=1$, respectively.  If two instances of the quantum  channel ${\map E}$ are combined through the quantum SWITCH,  the resulting channel is 
\begin{equation}
\mathcal{S}^{(2)}_{\omega_c}\left({\map E},{\map E} \right)[\rho_{B_1.. B_n}]=\mathcal{C}_+(\rho_{B_1.. B_n})\otimes\omega_c+\mathcal{C}_-(\rho_{B_1.. B_n})\otimes Z\omega_c Z,\label{A1}
\end{equation}
where,
\begin{eqnarray}
\mathcal{C}_+(\rho_{B_1B_2})&=& \left(\sum_{l=0}^3p_l^2q_l^2+p_0^2\left(1-q_0^2\right)+p_1^2\left(1-q_1^2\right)+p_2^2\left(1-q_2^2\right)+p_3^2\left(1-q_3^2\right)\right) \left[\rho_{B_1B_2}\right] \nonumber\\
&&+ 2q_0\sum_{l=0}^3p_l^2\left(q_1(\mathbf{I}\otimes X)[\rho_{B_1B_2}](\mathbf{I}\otimes X)+q_2(\mathbf{I}\otimes Y)[\rho_{B_1B_2}](\mathbf{I}\otimes Y) + q_3(\mathbf{I}\otimes Z)[\rho_{B_1B_2}](\mathbf{I}\otimes Z)\right) \nonumber \\
&&+ 2p_0\sum_{l=0}^3q_l^2\left(p_1(X\otimes\mathbf{I})[\rho_{B_1B_2}](X\otimes\mathbf{I})+p_2(Y\otimes\mathbf{I})[\rho_{B_1B_2}](Y\otimes\mathbf{I}) + p_3(Z\otimes\mathbf{I})[\rho_{B_1B_2}](\mathbf{I}\otimes Z)\right) \nonumber \\
&&+4(p_0p_1q_0q_1+p_2p_3q_2q_3)(X\otimes X)[\rho_{B_1B_2}] (X\otimes X)+4(p_0p_2q_0q_2+p_1p_3q_1q_3)(Y\otimes Y)[\rho_{B_1B_2}] (Y\otimes Y)\nonumber \\
&&+4(p_0p_3q_0q_3+p_2p_1q_2q_1)(Z\otimes Z)[\rho_{B_1B_2}] (Z\otimes Z)+ 4(p_0p_1q_0q_2+ p_2p_3q_1q_3)(X\otimes Y)[\rho_{B_1B_2}] (X\otimes Y) \nonumber \\
&&+ 4(p_0p_1q_0q_3+ p_2p_3q_1q_2)(X\otimes Z)[\rho_{B_1B_2}] (X\otimes Z) + 4(p_0p_2q_0q_1+ p_1p_3q_2q_3)(Y\otimes X)[\rho_{B_1B_2}] (Y\otimes X) \nonumber \\
&&+ 4(p_0p_2q_0q_3+ p_1p_3q_1q_2)(Y\otimes Z)[\rho_{B_1B_2}] (Y\otimes Z) + 4(p_0p_3q_0q_1+p_2p_1q_2q_3)(Z\otimes X)[\rho_{B_1B_2}] (Z\otimes X) \nonumber \\
&&+ 4(p_0p_3q_0q_2+p_1p_2q_1q_3)(Z\otimes Y)[\rho_{B_1B_2}] (Z\otimes Y);\label{A2}
\end{eqnarray}
\begin{eqnarray} 
\mathcal{C}_-(\rho_{B_1B_2})&=& ~~~2\sum_{l=0}^3p_l^2\left(q_1q_2(\mathbf{I}\otimes Z)[\rho_{B_1B_2}](\mathbf{I}\otimes Z)+q_2q_3(\mathbf{I}\otimes X)[\rho_{B_1B_2}](\mathbf{I}\otimes X)+ q_3q_1(\mathbf{I}\otimes Y)[\rho_{B_1B_2}](\mathbf{I}\otimes Y)\right) \nonumber \\
&&+2\sum_{l=0}^3q_l^2\left(p_1p_2(Z\otimes \mathbf{I})[\rho_{B_1B_2}](Z\otimes \mathbf{I})+p_2p_3(X\otimes \mathbf{I})[\rho_{B_1B_2}](X\otimes \mathbf{I})+ p_3p_1(Y\otimes \mathbf{I})[\rho_{B_1B_2}](Y\otimes \mathbf{I})\right) \nonumber \\
&&+ 4(p_2p_3q_0q_1 +p_0p_1q_2q_3)(X\otimes X)[\rho_{B_1B_2}] (X\otimes X) + 4(p_1p_3q_0q_2 +p_0p_2q_1q_3)(Y\otimes Y)[\rho_{B_1B_2}] (Y\otimes Y) \nonumber \\
&&+ 4(p_1p_2q_0q_3 +p_0p_3q_1q_2)(Z\otimes Z)[\rho_{B_1B_2}] (Z\otimes Z) + 4(p_0p_1q_1q_2 +p_2p_3q_0q_3)(X\otimes Z)[\rho_{B_1B_2}] (X\otimes Z) \nonumber \\
&&+ 4(p_0p_1q_1q_3 +p_2p_3q_0q_2)(X\otimes Y)[\rho_{B_1B_2}] (X\otimes Y) + 4(p_1p_3q_0q_1 +p_0p_2q_2q_3)(Y\otimes X)[\rho_{B_1B_2}] (Y\otimes X) \nonumber \\
&&+ 4(p_1p_3q_0q_3 +p_0p_2q_1q_2)(Y\otimes Z)[\rho_{B_1B_2}] (Y\otimes Z) + 4(p_1p_2q_0q_1 +p_0p_3q_2q_3)(Z\otimes X)[\rho_{B_1B_2}] (Z\otimes X) \nonumber \\
&&+ 4(p_1p_2q_0q_2 +p_0p_3q_1q_3)(Z\otimes Y)[\rho_{B_1B_2}] (Z\otimes Y).\label{A3}
\end{eqnarray}
For ${\map E}_1={\map E}_1={\map N}_{XY}$, putting $p_0=p_3=q_0=q_3=0$ and $p_1=p_2=q_1=q_2=\frac{1}{\sqrt{2}}$ in Eqs.(\ref{A2}-\ref{A3}) we obtain
\begin{eqnarray}
\mathcal{C}_+(\rho_{B_1B_2})&=&\rho_{B_1B_2}+Z\otimes Z(\rho_{B_1B_2})Z\otimes Z,\\
\mathcal{C}_-(\rho_{B_1B_2})&=&\mathbb{I}\otimes Z\left(\rho_{B_1B_2}\right)\mathbb{I}\otimes Z+Z\otimes \mathbb{I}\left(\rho_{B_1B_2}\right)Z\otimes \mathbb{I}.
\end{eqnarray}
Now considering $\omega_{c}=\ket{+}_c\bra{+}$, after proper normalization Eq.(\ref{A1}) becomes,
\begin{eqnarray}
\mathcal{S}^{(2)}_{\ket{+}_c}\left({\map E},{\map E}\right)[\rho_{B_1B_2}]&=&\frac{1}{4}\left[\mathbb{I}\otimes Z\left(\rho_{B_1B_2}\right)\mathbb{I}\otimes Z+Z\otimes \mathbb{I}\left(\rho_{B_1B_2}\right)Z\otimes \mathbb{I}\right]\otimes\mathbb{P}^{\ket{-}}_c\nonumber\\
&&+\frac{1}{4}\left[\rho_{B_1B_2}+Z\otimes Z\left(\rho_{B_1B_2}\right)Z\otimes Z\right]\otimes\mathbb{P}^{\ket{+}}_c,\label{A4}
\end{eqnarray}
where, $\mathbb{P}^{\ket{\pm}}:=\ket{\pm}\bra{\pm}$. Upon measuring the order system in $\{\mathbb{P}^{\ket{\pm}}\}$ basis, the conditional states of $B_1B_2$ read as,
\begin{eqnarray}
`-1' \mbox{outcome}&\rightarrow&\frac{1}{2}\left[\mathbb{I}\otimes Z\left(\rho_{B_1B_2}\right)\mathbb{I}\otimes Z+Z\otimes \mathbb{I}\left(\rho_{B_1B_2}\right)Z\otimes \mathbb{I}\right],\nonumber\\
`+1' \mbox{outcome}&\rightarrow&\frac{1}{2}\left[\rho_{B_1B_2}+Z\otimes Z\left(\rho_{B_1B_2}\right)Z\otimes Z\right].\label{A5}
\end{eqnarray}

\section{Random-receiver quantum communication for n receivers}\label{append}
When ${\map E}={\map F}:=\otimes_{k=1}^n{\map N}_{XY}$ the switched quantum distribution channel from Alice to $n$ Bobs read as, 
\begin{eqnarray}\label{NS1}
\mathcal{S}^{(n)}_{\omega_c}[\rho_{B_1.. B_n}]
&:=&\sum_{i_1,j_1,\cdots,i_n,j_n=0}^1 G_{i_1j_1.. i_nj_n}(\rho_{B_1.. B_n}\otimes\omega_c) G_{i_1j_1.. i_nj_n}^{\dagger};~~~~~~~~~\\
\mbox{where,}~~G_{i_1j_1.. i_nj_n}&:=& \otimes_{k=1}^nE_{i_k}F_{j_k}\otimes\ket{0}_c\bra{0}+ \otimes_{k=1}^nF_{j_k}E_{i_k}\otimes\ket{1}_c\bra{1};\nonumber\\
\mbox{with},~~E_0=F_0&=&\frac{1}{\sqrt{2}}X,~~\mbox{and},~~E_1=F_1=\frac{1}{\sqrt{2}}Y.\nonumber
\end{eqnarray}
Consider now an individual term in the summation of the right hand side of Eq.({\ref{NS1}}). The sign of the coherence term of the order qubit will be flipped , i.e., $\omega_{c}\to Z\omega_{c}Z$ if $\bigoplus_{k=1}^{n}(i_k\oplus j_k)=1$, and whenever $\bigoplus_{k=1}^{n}(i_k\oplus j_k)=0$ it will remain invariant, i.e., $\omega_{c}\to\omega_{c}$. Furthermore we will use the facts that $XX=YY=\mathbb{I}$ and $XY= i Z$ and $YX= -i Z$ in our following analysis.
	
{\bf Case-I}: Order bit invariant terms $\left[\bigoplus_{k=1}^{n}(i_k\oplus j_k)=0\right]$. In this case we have the following terms:
\begin{itemize}
\item[(i)] $i_k= j_k,~\forall~k;$ which will result the term $G_{i_1j_1.. i_nj_n}$ of the form $\mathbb{I}_{B_{1}...B_{n}}\otimes \left( |0\rangle_{c}\langle0|+ |1\rangle_{c}\langle1|\right) $ at right hand side of Eq.(\ref{NS1}).
\item[(ii)] For even number of cases (say $2m$) the indices $i_k$'s are different than the corresponding $j_k$'s and for the other cases they are equal. For a given $m\in\{0,1,...,[\frac{n}{2}]\}$ this will result terms $G_{i_1j_1.. i_nj_n}$ of the form $(-1)^m(Z_{1},...,Z_{2m},\mathbb{I}_{2m+1}...,\mathbb{I}_{n})\otimes\left( |0\rangle_{c}\langle0|+ |1\rangle_{c}\langle1|\right)$, where $(Z_{1},...,Z_{p},\mathbb{I}_{p+1}...,\mathbb{I}_{q})$ denotes term with $Z$ acting on $p$ among $q$ numbers of state and identity acting on rest. Number of such terms be ${q}\choose{p}$$ =\frac{q!}{p!(q-p)!}$.   
\end{itemize}
{\bf Case-II}: Order bit flipped terms $\left[\bigoplus_{k=1}^{n}(i_k\oplus j_k)=1\right]$. In this case we have the following terms:
\begin{itemize}
\item[(i)] For odd number of cases (say $2m+1$) the indices $i_k$'s and the corresponding $j_k$'s are $i_k=\bar{j}_k=0$ and for the other cases they are equal.. For a given $m\in\{0,1,...,[\frac{n-1}{2}]\}$ this will result terms $G_{i_1j_1.. i_nj_n}$ of the form $i\times(-1)^m(Z_{1},...,Z_{2m+1},\mathbb{I}_{2m+2}...,\mathbb{I}_{n})\otimes\left( |0\rangle_{c}\langle0|- |1\rangle_{c}\langle1|\right)$.
\item[(ii)]  $2m+1$ numbers of $i_k=\bar{j}_k=1$ and rests are equal. For a given $m\in\{0,1,...,[\frac{n-1}{2}]\}$ this will result terms $G_{i_1j_1.. i_nj_n}$ of the form $-i\times(-1)^m(Z_{1},...,Z_{2m+1},\mathbb{I}_{2m+2}...,\mathbb{I}_{n})\otimes\left( |0\rangle_{c}\langle0|- |1\rangle_{c}\langle1|\right)$.
\end{itemize}
	
Combining these all together we finally have,
\begin{eqnarray}\label{NS4}
\mathcal{S}^{(n)}_{\omega_c}[\rho_{B_1.. B_n}]
&:=&\frac{1}{2^n}\sum_{m=0}^{[\frac{n}{2}]}\left[ \left( Z_{1},...,Z_{2m},\mathbb{I}_{2m+1},...,\mathbb{I}_{n}\right) \rho_{B_{1}B_{2}...B_{n}}\left(  Z_{1},...,Z_{2m},\mathbb{I}_{2m+1},...,\mathbb{I}_{n}\right) \right] \otimes\omega_{c}\nonumber\\
&&+\frac{1}{2^n}\sum_{m=0}^{[\frac{n-1}{2}]}\left[ \left(  Z_{1},...,Z_{2m+1},\mathbb{I}_{2m+2},...,\mathbb{I}_{n}\right) \rho_{B_{1}B_{2}...B_{n}}\left( Z_{1},...,Z_{2m+1},\mathbb{I}_{2m+2},...,\mathbb{I}_{n}\right)\right]  \otimes Z\omega_{c}Z.
\end{eqnarray}
Suppose that initial state of order system is $\omega_c=\ket{+}\bra{+}$. In that case after the evolution of switched channel if depending on the outcome of Pauli $X$ measurement on order system if one of the receivers apply suitable local unitary correction on his subsystem, then the final outcome state reads as,
\begin{equation}
\frac{1}{2^n}\sum_{m=0}^{[\frac{n}{2}]}\left[ \left( Z_{1},...,Z_{2m},\mathbb{I}_{2m+1},...,\mathbb{I}_{n}\right) \rho_{B_{1}B_{2}...B_{n}}\left(  Z_{1},...,Z_{2m},\mathbb{I}_{2m+1},...,\mathbb{I}_{n}\right) \right].
\end{equation}
In the present context the input state is the generalized GHZ state $\ket\psi_{B_1\cdots B_n}=\alpha\ket{0\cdots0}_{B_1\cdots B_n}+\beta\ket{1\cdots1}_{B_1\cdots B_n}$ which is invariant under local unitary $Z$ operation by any even number of parties. Thus the state gets distributed perfectly among $n$ receivers. 

To reproduce the qubit information at one of the receivers' lab they will follow a LOCC protocol. Suppose that the qubit state needs to be reproduced at $i^{th}$ Bob. All other Bobs will perform Pauli $X$ measurement on their respective subsystems and inform the measurement results $x_k\in\{+1,-1\}$, $\forall~k\in\{1,\cdots,n\}~\&~k\neq i$. Depending on this information $i^{th}$ Bob will apply $Z$ unitary correction on his part if $\Pi_{k\neq i}x_k$ is $-1$, otherwise he does nothing.


\section{Random-receiver quantum communication with controlled operations in a definite order}\label{appene}

In this section we discuss the use of controlled operations in a definite causal order for random-receiver quantum communication. 
First, we show that, if arbitrary controlled operations are allowed, it is easy to construct protocols that achieve random-receiver quantum communication through entanglement-breaking channels. Then, we show that, if the controlled operations are restricted to be {\tt SWAP} operations, then random-receiver quantum communication cannot be achieved for any odd $n$.

\subsection{Protocol for random-receiver quantum communication using arbitrary controlled operations} 
The following protocol permits random-receiver quantum communication through the channels $\map A  =  \map B  =  \map N_{XY}^{\otimes n}$.    The protocol starts with the message encoded in the state of the first qubit.  First,   the sender applies a {\tt CNOT} gate to the message and the control qubit, initialized in the state $|+\>$.  As a result, the message and the control qubit end  up in the state $[|0\>  \, (  \alpha  |0\> + \beta |1\>)  +  |1\>  ( \alpha |1\>  + \beta  |0\>)]/\sqrt 2$.  Second, the sender prepares  $n-1$ qubits   in the state $|0\>$.  The sender sends the message and the other $n-1$ qubits  through the channel $\map A \map B  =(\map N_{XY}^{2})^{\otimes n}$, which collapses the overall state a classical mixture of the states $|0\>|0\>^{\otimes (n-1)}  \, (  \alpha  |0\> + \beta |1\>) $ and $|1\> |0\>^{\otimes (n-1)}  ( \alpha |1\>  + \beta  |0\>)$.  Third,  {\tt CNOT} gates are applied to the control qubit and to the additional $n-1$ qubits,    producing either the state $|0\>  (\alpha   |0\>^{\otimes n}  + \beta  |1\>^{\otimes n} )$ or the state $|1\>  \,  (\alpha |1\>^{\otimes n}    + \beta  |0\>^{\otimes n}   )$.   

Finally, the first receiver measures the first qubit in the computational basis, and, if the outcome is 1, all the other receivers perform   the bit flip operation $X  =  |0\>\<1|  +  |1\>\<0|$ on the remaining qubits and on the control qubit.  In this way, the remaining qubits  end up in the generalized GHZ state $\alpha |0\>^{\otimes n}   + \beta  |1\>^{\otimes n}$, and random-receiver quantum communication can be achieved as in the noiseless protocol presented in the introduction.  

Note that this protocol uses the control qubit to bypass the noisy channels $\map A$ and $\map B$, as one can see from the fact that,  after the channels $\map A$ and $\map B$ have acted,  all the information about the message is on the control qubit. 
  In addition, the protocol freely generates entanglement between the $n$ receivers after the channels $\map A$ and $\map B$ have acted. The entanglement generation is achieved by the $n-1$ {\tt CNOT} gates applied in the last step of the protocol. In general, these  {\tt CNOT}s  cannot be implemented by the receivers, due to their spatial separation.  Hence,  they must be regarded  as performed by a third party other than the receivers.   
  
Note that the presence of entangling operations between the control and each receiver is essential in the above protocol. More generally, entangling operations are necessary in any protocol that achieves random-receiver quantum communication through entanglement-breaking channels in a definite order.   Any such protocol needs to bypass the entanglement-breaking channels by encoding quantum information in the control qubit. An equivalent condition can be obtained by introducing an additional reference system at the receiver's end: in order to achieve perfect quantum communication, the protocol must transform a maximally entangled state of the input and the reference into a maximally entangled state of the control and the reference. After the action of an entanglement-breaking channel, the control and the reference have no correlation with the qubits at the receivers' locations.  Hence, no quantum information can be transferred back from the control to the receivers without the use of entangling operations.

\subsection{No GHZ state generation with controlled {\tt SWAP} operations}

Arbitrary controlled operations appear to be a too broad set for the problem of random-receiver quantum communication, in that they allow a complete transfer of information to the control qubit and violate the locality restrictions among the receivers.  On the other hand, 
The quantum SWITCH can be regarded as a controlled {\tt SWAP}   operation {\em in time}: it swaps the order of quantum systems appearing in a given time sequence, putting the inputs/outputs of channel $\map A$ either before or after the inputs/outputs of channel $\map B$.  One may then ask if controlled {\tt SWAP} operations {\em in space} can reproduce the same features when the channels $\map A$ and $\map B$ are arranged in a fixed sequential order, say with $\map A$ acting before $\map B$. 

\begin{figure}[h]
  \includegraphics[scale=0.5]{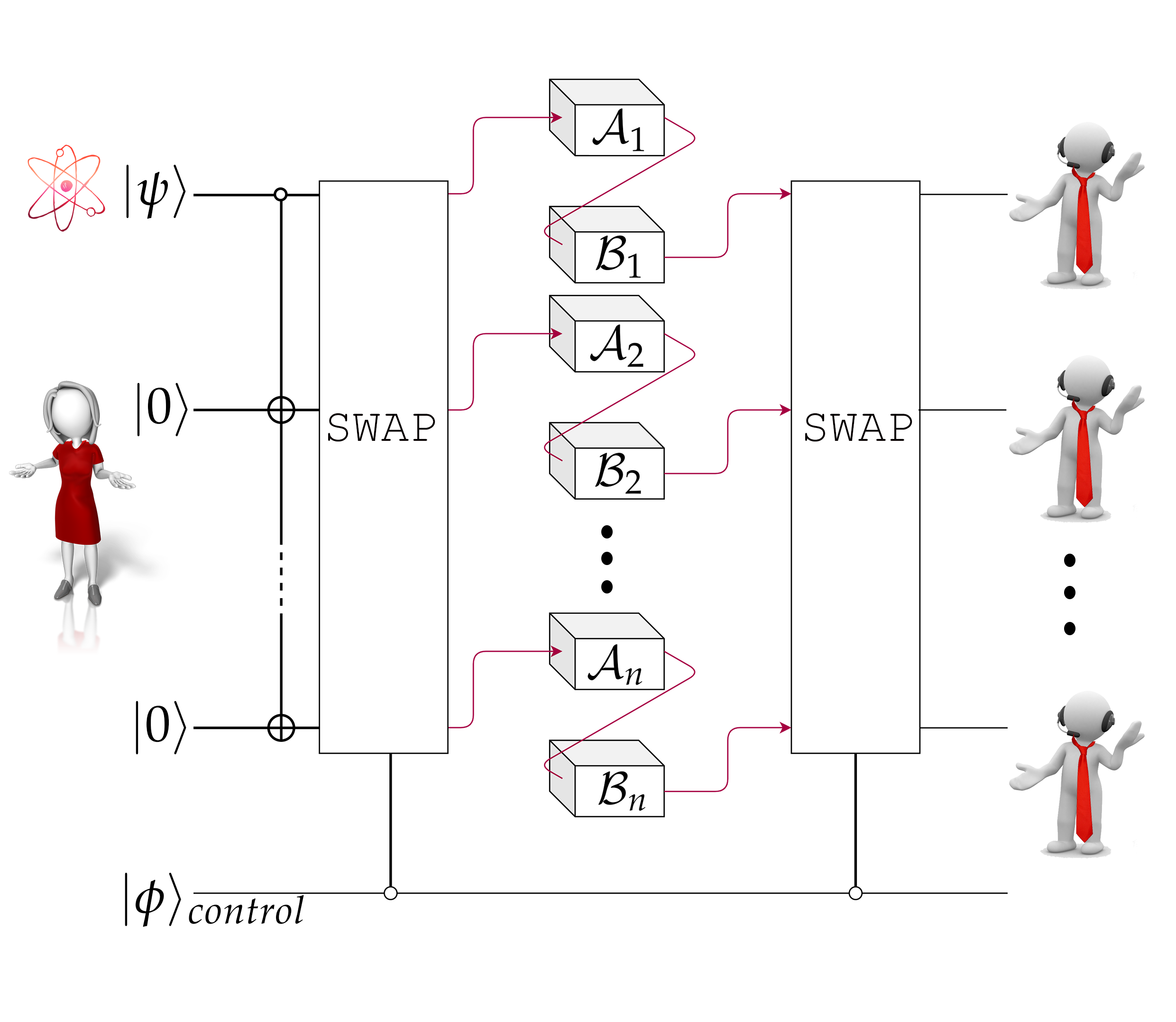}
  \caption{Random-receiver quantum communication task with controlled {\tt SWAP} operations {\em in space}.}
  \label{fig:swap}
\end{figure}

Here we show that the answer is negative, in the following sense: suppose that the quantum channels $\map A$ and $\map B$ are placed in a fixed order, and that their inputs and ouputs undergo controlled permutations, as in Figure \ref{fig:swap}.  For  odd $n$, we show that, no matter which controlled permutations are chosen, one of the receivers will remain in a fixed state, independent of the quantum message from the sender. 

The argument is simple. The output of channel  $\map N_{XY}^{\otimes n}$ is a mixture of states  in the computational basis, of the form  $\bigotimes_{j=1}^n  |b_j\>$, with $b_j  \in  \{0,1\}$ for every $j$.     
Perfect quantum communication through the entanglement-breaking channel  is possible only if, for every state appearing  in the mixture, the initial message has been perfectly transferred to the control qubit. 
 

Let us focus on one specific  state in the mixture, say $|\st b\>  = \bigotimes_{j=1}^n  |b_j\>$,   and let us denote by $|\psi'\>  = \alpha'|0\>  + \beta'|1\>$ the state of the control qubit conditional to the  state  $\bigotimes_{j=1}^n  |b_j\>$. 
   The state $  |\st b\>|\psi'\>$  then undergoes a controlled operation  $W$, becoming the state $W    |\st b \>|\psi'\>   =   \alpha'   \bigotimes_{j=1}^n  |b_{\pi(j)}\> \otimes |0\>   + \beta'   \bigotimes_{j=1}^n  |b_{\sigma(j)}\> \otimes |1\>$, where $\pi $ and $\sigma$ are two permutations.  Equivalently, the state can be rewritten as $W    |\st b \>|\psi'\>   =   \alpha'   \bigotimes_{j=1}^n  |b_{j}\> \otimes |0\>   + \beta'   \bigotimes_{j=1}^n  |b_{\tau(j)}\> \otimes |1\>$ for some suitable permutation $\tau$.  When $n$ is odd, there exists at least one value of $j$ such that $b_j  =  \tau(b_j)$. Hence, the $j$-th system ends up in the state $|b_j\>$, which has no dependence on the coefficients $\alpha'$ and $\beta'$, and therefore on the initial message.  
   
In summary, the $n$ output systems and the control end up in a mixture of pure states, one of which is a product between a fixed state of system $j$ and the remaining systems.  Hence, the quantum message cannot be transferred perfectly to the $j$-th receiver, since with some non-zero probability, the state of the $j$-th system will be independent of the message.

\end{document}